\documentclass[12pt]{article}\usepackage{a4}
\newcommand{\beao}{\begin{eqnarray*}}
\newcommand{\eeao}{\end{eqnarray*}}
\newcommand{\be}{\begin{equation}}\newcommand{\ee}{\end{equation}}
\newcommand{\bea}{\begin{eqnarray}}
\newcommand{\eea}{\end{eqnarray}}
\newcommand{\beq}{\begin{eqnarray}}
\newcommand{\eeq}{\end{eqnarray}}
\newcommand{\tra}{\top}
\newcommand{\nn}{\nonumber}
\newcommand{\pa}{\partial}

\newcommand{\al}{\alpha}\newcommand{\la}{\lambda}
\newcommand{\tr}{{\rm tr}}
\newcommand{\Ref}[1]{(\ref{#1})}
\newcommand{\sh}{\sinh(2s)}
\newcommand{\ch}{\cosh(2s)}
\newcommand{\p}{{p}}\newcommand{\q}{{q}}
\renewcommand{\u}{\tilde{u}}
\usepackage{epsfig}
\begin{document}
\title{Polarization tensor of charged gluons in color magnetic background field at finite temperature}
\author{M.~Bordag\thanks{e-mail: Michael.Bordag@itp.uni-leipzig.de}\\
{\small University of Leipzig, Institute for Theoretical Physics,} \\
{\small  Postfach 100 920, 04009  Leipzig, Germany} \\ [12pt]
{\small and} \\ [12pt]
V. Skalozub\thanks{e-mail: Skalozub@ff.dsu.dp.ua}\\
{\small Dnepropetrovsk National University, 49050 Dnepropetrovsk, Ukraine}}

\maketitle
\begin{abstract}
We calculate the polarization tensor of charged gluons in a
Abelian homogeneous magnetic background field at finite
temperature in one loop order Lorentz background field gauge in
full generality. Thereby we first determine the ten independent
tensor structures. For the calculation of the corresponding form
factors we use the Schwinger representation and represent form
factors as double parametric integrals and a sum resulting from
the Matsubara formalism used. The integrands are given explicitly
in terms of hyperbolic trigonometric functions. Like in the case
of  neutral gluons, the polarization tensor is not transversal.
Out of the tensor structures, seven   are transversal and three
are not. The nontransversal part follows explicitly from our
calculations.
\end{abstract}

\section{Introduction}
Investigations of non-Abelian gauge fields at finite temperature
which started   many years ago have been initiated  by two
fundamental phenomena, the electroweak  phase transition and the
deconfinement phase transition happening at high temperature.
Numerous methods of calculations were developed and applied and
important results on the  phase transitions and properties of the
fermion and gauge boson plasma have been obtained. The key element
of  most calculation schemes is the polarization tensor. It gives
information about the spectra of fields and it is necessary to
calculate various thermodynamic potentials beyond tree level. This
object was investigated  in both,  continuum quantum field theory
and lattice Monte Carlo calculations. In continuum theory, in
particular, there is the necessity of  resummations of
perturbation series because of infrared singularities appearing in
higher orders.  That was implemented in different resummation
schemes such as hard thermal loop resummations \cite{Blaizot02},
\cite{Braaten90a,Braaten92}, nonperturbative solution of the
Schwinger-Dyson equations \cite{Kalashnikov84}-\cite{Kraemmer95},
2PI effective action  \cite{Aarts03,Reinosa2007}, etc. A wide
bibliography on these problems is adduced in the review
\cite{Kraemmer04}.

In the past decade  it was realized     in both, continuum field
theory \cite{Starinets94,Enqvist:1994rm,Bordag00,Strelchenko04}
and Monte Carlo computations on a lattice
\cite{Agasian03,Demchik06} that at high temperature   magnetic
fields of the order $gB \sim g^4 T^2$, where $g$ is the gauge
coupling constant, $B$ is the field strength  $T$ is the
temperature, are spontaneously created. This phenomenon is
important for both high temperature QCD and  the early universe
where strong magnetic fields of different kinds had to be present
\cite{Giovannini:2003yn}. Hence, a comprehensive investigation of
non-Abelian gauge fields in a magnetic background at high
temperature is in order. Although such  investigations  also
started  many years ago, they  mainly dealt with the one-loop
calculations of the effective potential
\cite{Muller81}-\cite{Meisinger02} with the
 goal to study the influence of temperature on the
instability of the background field related to the tachyonic mode
  ($n = - 1$) which is present
in the charged gluon's spectrum $\epsilon^2_n = p^2_3 + (2 n + 1)
g B , n = -1, 0, 1, ....$, where $p_3 $ is the momentum component
along the field direction. At finite temperature, the situation
with the instability is different because a  temperature and field
dependent gluon magnetic mass can be generated and possibly remove
the instability from the spectrum. At present, this problem is not
finally settled although in some approximation it has been
considered in Refs. \cite{Meisinger02,Strelchenko00,Bordag00}. In
fact, only few attention was spent to study the radiation spectra
of gluons or $W$-bosons. Frequently, only fragments were
considered like the projection to the neutral sector or to tree
level states or with special gauge choices \cite{Strelchenko00},
\cite{Strelchenko05}. In addition, the question about the
transversality was not finally settled what caused a number of
problems over the years (see a recent discussion in
\cite{Bodeker:1999ud}).

In our previous papers \cite{Bordag:2005br,BORDAG2006} we
investigated the pure one-loop gluon polarization tensor in a
homogeneous magnetic background field, first at zero and later at
finite temperature. In \cite{Bordag:2005br} we calculated the
polarization tensor for both, the color neutral and the color
charged gluons, in covariant gauge at zero temperature. First we
determined  all possible tensor structure. It turned out that the
polarization tensor in the magnetic field is not transversal
(although the 'weak' transversality, i.e.,  if multiplied on both
side with the momentum, holds). The corresponding form factors
were represented within the Schwinger formalism as double
parametric integrals with explicitly known integrands. The
nontransversal structure was confirmed by an investigation of the
corresponding Slavnov-Taylor identities \cite{BORDAG2006}. In Ref.
\cite{BORDAG2006E} the neutral polarization tensor was calculated
at finite temperature. The additional tensor structures appearing
due to the temperature were determined and all form factors were
represented again in terms of the double parametric integrals and
an additional sum because of the temperature.  In the present
paper we make the final step in this series and calculate the
charged polarization tensor at finite temperature.

We are faced with two problems. First, we need to find all  tensor
structures $T_{\mu\nu}$ with the property that only $p_\mu
T_{\mu\nu} p_\nu=0$ holds where $p_\mu$ is a non-commuting
momentum because of the magnetic field (for details see below),
which to our knowledge had not yet been written down in a complete
form for nonzero temperature. The second problem is to represent
the form factors in a way expressing the weak transversality in an
explicit form. For this it was necessary to find in  the
parametric integrals the necessary structures which allowed to
integrate by parts. This is done and the remarkable property
observed earlier that the surface terms from the integration by
parts cancel just the contributions from the tadpole graphs up to
the Debye mass term is confirmed in this case too.

In the following we follow as close as  possible our previous
paper  \cite{BORDAG2006E}. In section 2 the basic notations are
introduced and   the tensor structures of the polarization tensor
are described. In section 3 the calculation of the form factors is
actually carried out. For instance, in section 3.5 the Debye mass
of charged gluons in the magnetic field
  is obtained. The Conclusion summarize the
main results, the explicit expressions for the one-loop form
factors and  some discussion. The procedure of restoring  the form
factors from the expressions calculated in
  the main text is given in the Appendix.

Throughout the paper we put $\hbar=c=k_B= g = 1 $ and  also the
magnetic field $B$ is put equal to one in the most part. The
dependence on $B$ can be restored by $T\to T/B^{1/2}$ for the
temperature and $\p_\mu\to p_\mu/B^{1/2}$ for the momenta.

\section{Basic notations}
We start from the operator structures $T^{(i)}_{\la\la'}$ which are allowed by the weak transversality condition,
\be\label{wtc}
p_\la \ T^{(i)}_{\la\la'} \ p_{\la'}=0 \ ,
\ee
which follows from the corresponding relation the polarization
tensor has to obey. Here $ p_\la$ is  the momentum depending on
the external field. These structures appear in the expansion in
terms of form factors which will be given below.
In the magnetic background field, the polarization tensor can be constructed out of the vectors $l_\mu$,  $h_\mu$ and $d_\mu$,
\bea&&\label{vecp} l_\mu=\left(\begin{array}{c}0\\0\\p_3\\p_4\end{array}\right),~
h_\mu=\left(\begin{array}{c}p_1\\p_2\\0\\0\end{array}\right), ~ d_\mu=\left(\begin{array}{c}p_2\\-p_1\\0\\0\end{array}\right),
\eea
where the third vector is  $d_\mu= F_{\mu\nu}p_\nu$ and we note
$p_\la=l_\la+h_\la$, and the matrixes
\be\label{F}
\delta^{||}_{\mu\lambda}=\left(\begin{array}{cccc}0&0&0&0\\0&0&0&0\\0&0&1&0
\\0&0&0&1\end{array}
\right), \
\delta^{\perp}_{\mu\lambda}=\left(\begin{array}{cccc}1&0&0&0\\0&1&0&0\\0&0&0&0
\\0&0&0&0\end{array}
\right), \
F_{\mu\lambda}=\left(\begin{array}{cccc}0&1&0&0\\-1&0&0&0\\0&0&0&0\\0&0&0&0\end{array}
\right).
\ee
Hence the operator structures can also be constructed out of these
quantities only. We mention that with our choice $B=1$, $F_{\mu\nu}$ in \Ref{F} is the field strength of the background field.

 Further,  the structures $T^{i}_{\la\la'}$ can be at most quadratic in  the momenta and following the polarization tensor they
  must be Hermite. Writing down all allowed combinations, a set of linear independent ones is
\begin{eqnarray} \label{Ti}
T^{(1)}_{\la\la'}&=& l^2 \delta^{||}_{\la\la'}-l_\la l_{\la'}\nn \\
T^{(2)}_{\la\la'}&=&h^2\delta^\perp_{\la\la'}+2iF_{\la\la'}
-h_\la h_{\la'}=d_\la d_{\la'} +iF_{\la\la'}   \nn \\
T^{(3)}_{\la\la'}&=& h^2 \delta^{||}_{\la\la'}+
l^2 \delta^\perp_{\la\la'} -l_\la h_{\la'}-h_\la l_{\la'}\nn \\
T^{(4)}_{\la\la'}&=& i(l_\la d_{\la'}-d_\la l_{\la'} )
+il^2F_{\la\la'}-\delta^{||}_{\la\la'}\nn \\
T^{(5)}_{\la\la'}&=&h^2 \delta^{||}_{\la\la'}-l^2 \delta^\perp_{\la\la'} \nn \\
T^{(6a)}_{\la\la'}&=&
\delta^{||}_{\la\la'}+l^2 iF_{\la\la'}, \qquad T^{(6b)}_{\la\la'}=
3\delta^{\perp}_{\la\la'}+h^2iF_{\la\la'},
\end{eqnarray}
where also the identity
$i(d_{\la}h_{\la'}-h_{\la}d_{\la'})=ih^2F_{\la\la'}+\delta^\perp_{\la\la'}$
was used.  In our previous paper \cite{Bordag:2005br} we mentioned
instead of the two structures $T^{(6a)}_{\la\la'}$ and
$T^{(6b)}_{\la\la'}$ only their sum,
\be\label{T6}
T^{(6)}_{\la\la'}=T^{(6a)}_{\la\la'}+T^{(6b)}_{\la\la'}.
\ee
In the end it will turn out that for the considered one-loop
contribution this is sufficient. However, for the calculations in
this section it is convenient to keep temporarily separately both,
$T^{(6a)}_{\la\la'}$ and $T^{(6b)}_{\la\la'}$.

Another remark on the properties of the structures
$T^{(i)}_{\la\la'}$ is  that the first four are transversal,
$p_\la T^{(i)}_{\la\la'}=T^{(i)}_{\la\la'}p_{\la'}=0$ holds for
$i=1,2,3,4$ in addition to \Ref{wtc}. The first three structures
are just a decomposition of the kernel of the quadratic part of
the action, Eq. (24) in \cite{Bordag:2005br},
\be\label{sumTipa}
T^{(1)}_{\la\la'}+T^{(2)}_{\la\la'}+T^{(3)}_{\la\la'}=K_{\la\la'}(p).
\ee

In the case of finite temperature, in addition to the magnetic
field,  we have to account also for the vector $u_\la$ which
describes the speed of the heat bath. In the following we assume
it to be orthogonal to $h_\la$. In fact we use $u_\la=(0,0,0,1)$.
With this vector additional tensor structures obeying the weak
transversality condition \Ref{wtc} can be constructed,
\bea\label{Tt}
T^7_{\la\la'}&=&
    \left(u_{\la}l_{\la'}+l_{\la}u_{\la'}\right)  (up)
    -\delta^{||}_{\la\la'}(up)^2-u_{\la}u_{\la'} l^2 \ ,
\nn \\
T^8_{\la\la'}&=&
 \left(u_{\la}h_{\la'}+h_{\la}u_{\la'} \right)  (up)
    -\delta^{\perp}_{\la\la'}(up)^2-u_{\la}u_{\la'} h^2    \ ,
\nn \\
T^9_{\la\la'}&=&
      u_{\la}id_{\la'}-id_{\la}u_{\la'} +2iF_{\la\la'}(up)   \ ,
\nn \\
T^{10a}_{\la\la'}&=&
      \delta^{||}_{\la\la'}(up)^2-u_{\la}u_{\la'} l^2, \quad
      T^{10b}_{\la\la'}=
      \delta^{\perp}_{\la\la'}(up)^2-u_{\la}u_{\la'} h^2 \ .
\eea
These structures are linear independent. Below, however, it will
turn out that $T^{10a}$ and $T^{10b}$ appear at the one-loop level
which we consider here only in intermediate steps and drop out
from the final result.

In addition to \Ref{Tt} there exists a further structure,
\be
\label{TD}
T^{\rm D}_{\la\la'}= u_{\la}u_{\la'}
\ee
which fulfills \Ref{wtc} for $p_4=0$ (in an obvious, trivial way).
Having in mind that the condition \Ref{wtc} holds only if the
external moment p of the polarization tensor fits into the
Matsubara formalism, i.e., if it is given by $p_4=2\pi Tl$ ($l$
integer) than \Ref{TD} makes sense as a structure being present
for $l=0$ only (or, formally, being proportional to
$\delta_{l,0}$). It is just this structure which delivers the
Debye mass term. It must be mentioned that $T^{\rm D}_{\la\la'}$
is contained as special case also in $T^8_{\la\la'}$,
$T^9_{\la\la'}$,  $T^{10a}_{\la\la'}$ or in $T^{10b}_{\la\la'}$
but due to its exceptional role it makes sense to keep it as
separate contribution.

Before writing down the expansion of the polarization tensor in
terms of form factors we mention one property which follows
directly from the basic commutator relation the momentum $p_\mu$
obeys, namely $p_\la
p^2=(p^2\delta_{\la\la'}+2iF_{\la\la'})p_{\la'}$. As a
consequence, for a function of $p^2$ the relations
\bea\label{p2+2iF}
p_\la f(p^2)&=&f(p^2+2iF)_{\la\la'}p_{\la'} \ ,
\nn \\
f(p^2)p_\la &=&p_{\la'}f(p^2+2iF)_{\la'\la}
\eea
hold where now $f$ must be viewed as a function of a matrix so
that it  itself becomes a matrix carrying the indices $\la$ and
$\la'$. The same is true with $h^2$ in place of $p^2$.

The decomposition of the polarization tensor can be written in the form
\be\label{expp} \Pi_{\la\la'}(p)=\sum_{i}\ \Pi^{(i)}(l^2,h^2+2iF)_{\la\la''} \  T^{(i)}_{\la''\la'}
+\Pi^{\rm D} T^{\rm D}_{\la\la'}.
\ee
The sum includes in general all structures $T^{(i)}_{\la\la'}$
defined in  \Ref{Ti} and in \Ref{Tt}. The form factors
$\Pi^{(i)}(l^2,h^2+2iF)_{\la\la'}$ depend on
$l^2$ and $h^2$ only (besides their dependence on the matrices in
\Ref{F}).  In \Ref{expp} the form factors can be placed also on
the right from the operator structures   applying both relations
\Ref{p2+2iF}.

\section{Calculation of the polarization tensor}
The basic Feynman graph for the polarization tensor  is shown in
Fig.5 in \cite{Bordag:2005br} and the notations of the vertex factors in
Fig.2 there. The analytic expression in momentum space reads
\bea\label{CPT}
\Pi_{\la\la'}(p)&=&\int\frac{dk}{(2\pi)^4} \ \left\{
\Gamma_{\la\nu\rho}G_{\nu\nu'}(p-k)\Gamma_{\la'\nu'\rho'}G_{\rho\rho'}(k)
\right.  \nn \\ && \left.
+(p-k)_{\la}G(p-k)k_{\la'}G(k)+k_{\la}G(p-k)(p-k)_{\la'}G(k)
\right\}  \nn \\ [6pt]&&
+\Pi^{\rm tadpol}_{\la\la'} \ ,
\eea
where the second line results from the ghost contribution
and the tadpole contribution is given by
\bea\label{tadpol}
\Pi^{\rm tadpol}_{\la\la'}&=&
\int\frac{dk}{(2\pi)^4} \ \left\{\delta_{\la\la'}G_{\rho\rho}(k)-G(k)_{\la\la'}
\right\} \nn \\ &&
+\int\frac{dp'}{(2\pi)^4} \ \left\{
\delta_{\la\la'}G_{\rho\rho}(p')+G_{\la'\la}(p')-2G_{\la\la'}(p')\right\}.
\eea
The vertex factor is given by
\be\label{Vf}
\Gamma_{\la\nu\rho}=
(k-2p)_\rho  \ \delta_{\la\nu}+\delta_{\rho\nu}(p-2k)_\la+\delta_{\rho\la}(p+k)_\nu \ .
\ee
For the following we rearrange it in the form
\be\begin{array}{rcllll}
\label{vf1}\Gamma_{\la\nu\rho}&=&
\underbrace{(p-2k)_\la  \ \delta_{\nu\rho}}&
+\underbrace{2\left(p_\nu\delta_{\la\rho}-p_\rho\delta_{\la\nu}\right)}&
+\underbrace{\left(-(p-k)_\nu\delta_{\la\rho}+k_\rho\delta_{\la\nu}\right)}
, \\ [8pt]
&\equiv & \ \ \ \ \ \ \ \Gamma^{(1)}_\la  \ \ \ \ \ &
+ \ \ \ \ \ \ \ \ \ \ \Gamma^{(2)}_{\la\nu\rho}&+\ \ \ \ \ \ \ \ \ \ \ \ \ \ \ \Gamma^{(3)}_{\la\nu\rho},
\end{array}
\ee
where $\Gamma^{(3)}_{\la\nu\rho}$ will be temporary further subdivided into two parts,
\be\label{vf2}\Gamma^{(3_1)}_{\la\nu\rho}=-(p-k)_\nu\delta_{\la\rho}
\quad \mbox{and}  \quad
\Gamma^{(3_2)}_{\la\nu\rho}=k_\rho\delta_{\la\nu} \ .
\ee

The momentum integration in the polarization tensor is carried out
using the formalism   introduced by Schwinger, \cite{Schwinger:1973kp}.
There the propagators (in Feynman  gauge, $\xi=1$) are represented
as parametric integrals,
\be
\label{prop1}G(p-k) =\int_0^\infty ds \ e^{-s(p-k)^2} \mbox{ , }\quad  G(k)=\int_0^\infty dt \ e^{-tk^2}
\ee
for the scalar lines and
\bea\label{prop2}
G_{\nu \nu'} (p-k)&=&\int_0^\infty ds \ e^{-s(p-k)^2}E_{\nu\nu'} \nn \\ 
G_{\rho \rho'} (k)&=&\int_0^\infty dt \ e^{-tk^2}\delta_{\rho\rho'} \eea
for the vector lines with
\be
\label{E0}E_{\nu\nu'}\equiv e^{-2isF} =\delta^{||}_{\la\la'}-iF_{\la\la'}\sinh(2s)+\delta^\perp_{\la\la'} \cosh(2s).
\ee
In this formalism the momentum integration is written as an
averaging procedure in some auxiliary space and for the basic
exponential
\be\label{Theta1} \hat{\Theta}=e^{-s(p-k)^2}e^{-tk^2}
\ee
it holds
\be\label{Av}
\int\frac{dk}{(2\pi)^4} \ \hat{\Theta} = \langle \hat{\Theta}\rangle =\Theta(l^2,h^2)
\ee
with
\be\label{Theta2}\Theta(l^2,h^2)=
\frac{\exp\left[-H\right]}{(4\pi)^2(s+t)\sqrt{\Delta}},
\ee
which is the result of the corresponding calculations, see
\cite{Schwinger:1973kp} for details. The following notations are used,
\be\label{H}H=\frac{st}{s+t}l^2+m(s,t)h^2
\ee
and
\bea\label{mst}
m(s,t)&\equiv& s-{\rm arctanh}\frac{p}{q}
\nn \\ &=&\frac12 \ln\frac{1+2t-e^{-2s}}{1-(1-2t)e^{-2s}}
\eea
as well as
\bea\label{N}\Delta&=&\left(q^2-p^2\right)/4 \nn \\&=&
t^2+t \sh+p/2
\eea
with the notations
\bea\label{pq}p&=&\ch-1 \ , \\q&=&2t+\sh\ ,
\eea
which will be met frequently in the following.
With these notations the self energy graph for scalar lines becomes represented by the parametric integrals in the form
\be
\label{SK}
\Pi^{\rm scalar}_{(T=0)}=\int\frac{dk}{(2\pi)^4} \ G(p-k)G(k)=\int dsdt \ \Theta(l^2,h^2).
\ee

These formulas are derived for $T=0$. To include nonzero
temperature, within the  Matsubara formalism we are using we have
to substitute the integration over the continuous momentum $k_4$
by a discrete sum over $l$ in $k_4=2\pi lT$,
\be\label{k4T}\int_{-\infty}^\infty\frac{dk_4}{2\pi} \ \to \ T\sum_{l=-\infty}^\infty \ .
\ee
In order to incorporate this into into the parametric integral we represent
\bea\label{}T\sum_{l=-\infty}^\infty \ &=&T\sum_{l=-\infty}^\infty \ \int_{-\infty}^\infty dk_4 \ \delta(k_4-2\pi lT)
\nn \\ &=&
T\sum_{l=-\infty}^\infty \ \int_{-\infty}^\infty d\sigma \
e^{-i\sigma 2\pi lT} \int_{-\infty}^\infty\frac{dk_4}{2\pi} \ e^{i\sigma k_4}
\eea
in this way keeping the original formalism on the expense of
accommodating the  additional factor $\exp(i\sigma k_4)$ and
carrying out the integration over $\sigma$ and the summation over
$l$ afterwards. Under the above assumption that the speed of the
heat bath is orthogonal to $h_\la$, the additional factor
$\exp(i\sigma k_4)$  can be incorporated into Schwinger's
formalism quite trivially because the integration over $k_4$ (in
the same way as that over $k_3$) decouples from the other ones in
the sense that the corresponding integrals factorize. For the
integration over $k_4$ we have
\be\label{k4la1}
\int_{-\infty}^\infty \frac{dk_4 }{2\pi} \
                            e^{i\sigma k_4} \ e^{-s(p_4-k_4)^2-tk_4^2} \
=\frac{\exp\left(-\frac{\sigma^2}{4(s+t)}+i\frac{\sigma sp_4}{s+t}-\frac{st}{s+t}p_4^2\right)}{\sqrt{4\pi(s+t)}}
\ee
and
\bea\label{k4la2}
&&T\sum_{l=-\infty}^\infty \ \int_{-\infty}^\infty d\sigma \
\int_{-\infty}^\infty \frac{dk_4 }{2\pi} \
                            e^{-\sigma k_4} \ e^{-s(p_4-k_4)^2-tk_4^2} \
\\ \nn&&~~~~~~~~~~~~~~~~~~~~~~~
=T\sum_{l=-\infty}^\infty \
    \exp\left(-(2\pi lT)^2+4\pi lTs (up)-\frac{s^2}{s+t}(up)^2
        -\frac{st}{s+t}p_4^2\right)    .
\eea
In this formula we used $(up)=p_4$ because this form of writing will be more convenient below.

Combining these with \Ref{Av}, \Ref{Theta2} and \Ref{Av} we get for a graph consisting of scalar lines
\bea\label{skT}
\Pi^{\rm scalar}&\equiv&
T\sum_{l=-\infty}^\infty \int\frac{d^3k}{(2\pi)^3} \ G(p-k)G(k)
\\ \nn&
=& \ T\sum_{l=-\infty}^\infty \ \int dsdt\
e^{-(2\pi lT)^2(s+t)+4\pi lTs (up)-\frac{s^2}{s+t}(up)^2}
 \sqrt{4\pi(s+t)}   \    \Theta(l^2,h^2) \ .
\eea
This representation is still not in a form which is sufficiently
convenient  for the following, for instance, it contains still the
ultraviolet divergence which appears from small $s$, $t$ and large
$l$. Using the well known resummation formula
\be\label{resum}
\sum_l\exp\left(-zl^2+al\right)=
\sqrt{\frac{\pi}{z}}\sum_N\exp\left(-\frac{\pi^2N^2}{z}
                        +i\pi N\frac{a}{z}+\frac{a^2}{4z^2}\right)
\ee
(both sums run over the integers) we obtain with $z\to (2\pi T)^2$, $a\to 4\pi T(up)$
\be\label{Psc}
\Pi^{\rm scalar}=\sum_N \int dsdt \ \Theta_T(l^2,h^2)\ .
\ee
Here we introduced the basic average
\be\label{ThetaT}
\Theta_T(l^2,h^2)= \ \exp\left\{-\frac{N^2}{4(s+t)T^2}+2s(\tilde{u}p)\right\}\Theta(l^2,h^2)\ ,
\ee
where $\Theta(l^2,h^2)$ given by Eq.\Ref{Theta2}  and  where the notation
\be\label{uti}\tilde{u}_\la=\frac{iN}{2(s+t)T}\ u_\la
\ee
was introduced.
This average is what comes at finite temperature in place of \Ref{Av},
\be
\label{AvT}
T\sum_{l=-\infty}^\infty \int\frac{d^3k}{(2\pi)^3} \
=
 \langle \hat{\Theta}\rangle_T =\Theta_T(l^2,h^2) \ .
\ee

In \Ref{ThetaT} the ultraviolet divergence is in the $N=0$
contribution taken  at $B=0$. The well known basic properties of
the representations as sum over $l$ or as sum over $N$ are that
the sum over $l$ is convenient for high temperature ($l=0$ -- the
so called static mode, gives the leading contribution for
$T\to\infty$) and the sum over $N$ gives the low temperature
expansion, for instance the $N=0$ term is the contribution at
$T=0$.

At this place it is meaningful to show how the magnetic field can be restored in the expressions involving the parameters $s$ and $t$.
Since in the proper
time representation of the Green's functions \Ref{prop1}, \Ref{prop2}
the phase is dimensionless and since the restoration of the $B$-dependence has to respect that, we must take $s(p - k)^{2} \to (sB)((p - k)^{2})/B$. Hence in the parameters $s$ and $t$ the magnetic field is restored by $s \to s B$ and  $t \to t B$.

We continue with the remark that since below $\Theta(l^2,h^2)$
will become part of the form factors it  is meaningful to write it
as a function of $h^2+2iF$. This can be done by means of the
relation
\be\label{ThetaZ}\Theta(l^2,h^2)=\Theta(l^2,h^2+2iF)Z
\ee
with
\be\label{Z}Z=-E^{\tra}\frac{D}{D^\tra}=
\delta^{||}+\frac{\al}{4N}iF+\frac{\beta}{4N}\ \delta^\perp \ ,
\ee
where the notations
\bea\label{albe}
\al  &=&\left(p^2+q^2\right)\sh-2pq\ch ,\nn \\
\beta&=&\left(p^2+q^2\right)\ch-2pq\sh ,
\eea
and
\be\label{AD}A=E-1 \ , \qquad   D=A-2itF
\ee
are introduced.

From Schwinger's formalism we need also the commutator relation
\be\label{p(s)} p_\mu \Theta(l^2,h^2)=\Theta(l^2,h^2)\overline{p}_\mu
\ee
with
\be\label{p(s)1}\overline{p}_\mu=\left(Ep\right)_\mu-\left(Ak\right)_\mu.
\ee
Here we used obvious short notations like
$\left(Ep\right)_\mu=E_{\mu\mu'}p_{\mu'}$.  Finally, we need the
average formulas for vectors, \bea
\label{Av1}\langle \hat{\Theta}k_\la\rangle_T &=&\langle  \hat{\Theta}\rangle_T \left(\frac{A}{D}p + \tilde{u} \right)_{\la}, \nn \\
\label{Av2}\langle \hat{\Theta}k_\la k_{\la'}\rangle_T &=&
\langle \Theta\rangle_T \left[
\left(\frac{A}{D}p+ \tilde{u} \right)_\la
\left(\frac{A}{D}p+ \tilde{u} \right)_{\la'}
+\left(\frac{iF}{D^\top}\right)_{\la\la'}\right],
\eea
together with the explicit representation
\be\label{A/D}\frac{A}{D}=\frac{s}{s+t}\ \delta^{||}- \ \frac{tp}{2N}   \ iF +\frac{p+t\sh}{2N} \ \delta^\perp.
\ee
All other quantities can be calculated from these, for example,
\be\label{umr}
\frac{iF}{D^\tra}=\frac12 \ \left(\frac{-2iF}{D}\right)^\tra=
\frac{1}{2t}\left(1-\frac{A}{D}\right)^\tra=\frac12\left(\frac{\delta^{||}}{s+t}+\frac{p}{2N}\ iF+\frac{q}{2N}\ \delta^\perp\right) \ .
\ee
Perhaps it is useful to remark that all these matrices commute one
with  the other, that the transposition changes the sign of $F$
and that the simple algebra ${\delta^{||}}^2=\delta^{||}$,
${\delta^{\perp}}^2=\delta^{\perp}$, $F^2=-\delta^{\perp}$,
$\delta^{||}\delta^\perp=\delta^{||}F=0$ and $\delta^{\perp}F=F$
holds.

The averages in \Ref{Av2} are calculated at $T\ne 0$. For $T=0$
they reduce to  the formulas known from \cite{Schwinger:1973kp}.
For $T\ne 0$ one needs to consider the corresponding
generalizations of Eq.\Ref{k4la1},
\be\label{}\int_{-\infty}^\infty \frac{dk_4 }{2\pi} \ k_4 \
                            e^{i\sigma k_4} \ e^{-s(p_4-k_4)^2-tk_4^2}\ .
\ee
Replacing the additional factor $k_4$ by $i\frac{\pa}{\pa
\sigma}$, after  integration over $\sigma$ an additional factor
$2\pi lT$ appears in a formula which generalized Eq.\Ref{skT}. It
remains to do the resummation from $l$ to $N$. Taking the
derivative with respect to $a$ from Eq.\Ref{resum} after some
calculations the first line in \Ref{Av2} for $\la=4$ appears. The
derivation of \Ref{Av2} is then finished by the remark that for
$\la=1,2,3$ no additional contributions appear. In a similar way
also the second line in Eq.\Ref{Av2} can be derived.

Now we turn to the calculation of the polarization tensor
\Ref{CPT}. Using   \Ref{prop2}, \Ref{Theta1}, \Ref{Av} and
\Ref{p(s)} it can be written in the form
\bea\label{}
\Pi_{\la\la'}&=&\sum_N \int dsdt \ \langle\hat{\Theta_T}\left[
\Gamma_{\la\nu\rho}E_{\nu\nu'}\Gamma_{\la'\nu'\rho}
+(\overline{p}-k)_{\la}k_{\la'}+k_{\la} (p-k)_{\la'}\right]\rangle \nn \\
&+& \ \Pi^{\rm tadpol}_{\la\la'} \ ,
\eea
where in $\Gamma_{\la\nu\rho}$ one needs to substitute $p$ by $\overline{p}$.

At the next step we divide the whole expression into parts according to the division made in \Ref{vf1} and \Ref{vf2},
\be\label{Pi2}
\Pi_{\la\la'}=
\sum_N \int dsdt \ \langle\hat{\Theta_T}\left[ \sum_{i,j}\hat{M}^{i,j}_{\la\la'}+
                                \hat{M}^{\rm gh}_{\la\la'}\right]\rangle
+ \ \Pi^{\rm tadpol}_{\la\la'}
\ee
with
\be\label{}\hat{M}^{i,j}_{\la\la'}=
\Gamma_{\la\nu\rho}^{(i)}E_{\nu\nu'}\Gamma_{\la'\nu'\rho}^{(j)}
\ee
and $\hat{M}^{\rm gh}_{\la\la'}$ is the corresponding contribution from the ghost loop.

The sums in \Ref{Pi2} include also the decomposition \Ref{vf2}.
The explicit expressions for these quantities read
\be\label{M1}
\begin{array}{lcl}
\hat{M}^{1,1}&=&
    \left(\overline{p}-2k\right)_\la\left(p-2k\right)_{\la'}\ \tr E,
\\
\hat{M}^{1,2}&=&2\left(\overline{p}-
    2k\right)_\la\left(\left(E-E^\tra\right)p\right)_{\la'} ,
\\
\hat{M}^{1,3_1}&=&-\left(\overline{p}-
    2k\right)_\la\left(E\left(p-k\right)\right)_{\la'},
\\
\hat{M}^{1,3_2}&=&Z\left(\overline{p}-
    2k\right)_\la\left(E^\tra k\right)_{\la'},
\\
\hat{M}^{2,1}&=&2\left(\left(E^\tra-E\right)\overline{p}\right)_\la (p-2k)_{\la'},
\\
\hat{M}^{2,2}&=&4\delta_{\la\la'}\left(\overline{p}Ep\right)-
    4\left(E^\tra\overline{p}\right)_{\la'}p_\la
    \\&&
    +4\left(Z\right)_{\la\la'}\left(\overline{p}p\right)
    -4\left(\overline{p}\right)_{\la'}\left(Ep\right)_\la ,
\\
\hat{M}^{2,3_1}&=&-2\delta_{\la\la'}\left(\overline{p}E(p-k)\right)+
    2\overline{p}_{\la'}\left(E(p-k)\right)_\la ,
\\
\hat{M}^{2,3_2}&=&-2\left(E\right)_{\la\la'}\left(\overline{p}k\right)+
    2\left(E^\tra\overline{p}\right)_{\la'}k_\la,
\\
\hat{M}^{3_1,1}&=&-\left(E^\tra \left(\overline{p}-
    k\right)\right)_\la(p-2k)_{\la'},
\\
\hat{M}^{3_2,1}&=&\left(Ek\right)_\la(p-2k)_{\la'},
\\
\hat{M}^{3_1,2}&=&-2\delta_{\la\la'}\left(\left(\overline{p}-k\right)Ep\right)
    +2\left(E^\tra\left(\overline{p}-k\right)\right)_{\la'}p_\la,
\\
\hat{M}^{3_2,2}&=&-2\left(E\right)_{\la\la'}(kp)+2k_{\la'}\left(Ep\right)_\la,
\\
\hat{M}^{\rm 3,3}&=&
\delta_{\la\la'}\left(\left(\overline{p}-k\right)E(p-k)\right)+E_{\la\la'}k^2
    -k_\la\left(E^\tra\left(\overline{p}-k\right)\right)_{\la'}
    -\left(E\left(p-k\right)\right)_\la k_{\la'}
\nn\\
\hat{M}^{\rm gh}&=&\left(\overline{p}-k\right)_\la k_{\la'}+
   k_\la(p-k)_{\la'}.
\end{array}
\ee
Using Eq. \Ref{p(s)1} and   simple relations like $A+2=E+1$  and
$EE^\tra=1$ these expressions can be simplified further.
After that we apply the average formulas \Ref{Av2} and  pass from
$\Theta(l^2,h^2)$ to  $\Theta(l^2,h^2+2iF)$ by means of
\Ref{ThetaZ} which brings a factor $Z$ to the  $M^{i,j}$ which
couples to the index $\la$, in detail, $Z M^{i,j}$ stands for
$Z_{\la\la''}M^{i,j}_{\la''\la'}$. In this way we come to
\be\label{Mij}
\langle \hat{\Theta}\hat{M}^{i,j}\rangle_T =
    \Theta_T(l^2,h^2+2iF)\ M^{i,j}
\ee
with
\bea\label{M3}
M^{1,1}&=&\left(\left(Pp-2\u\right)_\la \left(P^\tra p-2\u\right)_{\la'}+
    2Z\frac{2iF}{D^\tra}\right)\ \tr E
\nn \\
M^{1,2}&=& \left(Pp-2\u\right)_\la \left(Rp\right)_{\la'},
\nn \\
M^{1,3_1}&=&\left(Pp-2\u\right)_\la \left(-Q_{31}p+\u\right)_{\la'}
                                            -ZE^\tra\frac{2iF}{D^\tra}
\nn \\
M^{1,3_2}&=&
    \left(Pp-2\u\right)_\la \left(Q_{32}p+\u\right)_{\la'}
    -ZE\frac{2iF}{D^\tra}
\nn \\
M^{2,1}&=&\left(R^\tra p\right)_\la \left(P^\tra p-2\u\right)_{\la'} \ ,
\nn \\
M^{2,2}&=&-4\left[
    \left(Sp\right)_\la \left(Tp\right)_{\la'}+
    \left(T^\tra p\right)_\la \left(S^\tra p\right)_{\la'}
    -S_{\la\la'}\left(pT^\tra p\right)  -T^\tra_{\la\la'}\left(pSp\right)\right]
     \nn \\&&   +4ST^\tra 2iF \ ,
\nn \\
M^{2,3_1}&=&2
    \left(V^\tra p-\u\right)_\la
    \left(S^\tra p\right)_{\la'}
    -2Z_{\la\la'}\left(\left(p\frac{2itF}{D^\tra}p\right)-(\u p)\right)
                         -2iFZ\frac{2itF}{D^\tra} \ ,
\nn \\
M^{2,3_2}&=&
    2\left(Up+\u\right)_\la \left(Tp\right)_{\la'}
    -2\left(EZ\right)_{\la\la'}
        \left(\left(p\left(\frac{A}{D}\right)^\tra p\right)+(\u p)\right)
    +ZA\frac{2iF}{D^\tra}\ ,
\nn\\
M^{3_1,1}&=&\left(-Q_{31}^\tra p+\u\right)_{\la}\left(P^\tra p -2\u\right)_{\la'}
                                            -ZE^\tra\frac{2iF}{D^\tra} \ ,
 \\
M^{3_2,1}&=&
    \left(Q_{32}^\tra p+\u\right)_{\la}\left(P^\tra p-2\u\right)_{\la'}
    -ZE\frac{2iF}{D^\tra} \ ,
\nn \\
M^{3_1,2}&=&
    2\left(Sp\right)_\la \left(Vp-\u\right)_{\la'}
    -2Z_{\la\la'}\left(\left(p\frac{2itF}{D^\tra}p\right)-(\u p)\right)
    -2iFZ\frac{2itF}{D^\tra} \ ,
\nn \\
M^{3_2,2}&=&
    2\left(T^\tra p\right)_\la \left(U^\tra p+\u\right) _{\la'}
    -2\left(EZ\right)_{\la\la'}
            \left(\left(p\left(\frac{A}{D}\right)^\tra p\right)+(\u p)\right)
    +ZA\frac{2iF}{D^\tra} \ ,
\nn \\
M^{3,3}+M^{\rm gh}&=&
    Z_{\la\la} \left(
    \left(\left(    \left(1-\frac{A}{D}\right)p-\u\right)
                    \left(\left(1-\frac{A}{D}\right) p-\u\right)\right)
            -\tr E\frac{iF}{D^\tra} \ \right)
\nn \\  && + \left(EZ\right)_{\la\la'}  \left(
                    \left(\frac{A}{D}p+\u\right)\left( \frac{A}{D} \ p+\u\right)
            +\tr \frac{iF}{D^\tra} \ \right)  \ , \nn
\eea
where in the last line some cancellations occurred.
Here, again, new notations were introduces, namely
\bea\label{PQetc}
P&=&\left(1-2\frac{A}{D}\right)^\tra=Z\left(E-(E+1)\frac{A}{D}\right),\nn \\
&=&-\frac{s-t}{s+t}\ \delta^{||}-\frac{tp}{\Delta}\ iF-\frac{p/2-t^2}{\Delta}\ \delta^\perp
\equiv r_3 \, \delta^{||}+\al_3 \, iF+\beta_3 \, \delta^\perp,
\nn \\ [10pt]
R&=&2\left(E-E^\tra\right)=-4\sh \ iF \ ,
\nn \\ [10pt]
Q_{31}&=&\left(Z\left(1-\frac{A}{D}\right)\right)^\tra = E\frac{-2itF}{D},\nn \\
&=&\frac{t}{s+t}\ \delta^{||}+t\frac{p\ch-q\sh}{2\Delta}\ iF
                                        +t\frac{q\ch-p\sh}{2\Delta}\ \delta^\perp,
\nn \\ &  \equiv &s_4 \, \delta^{||}+\gamma_4 \, iF+\delta_4 \, \delta^\perp,
\nn \\ [10pt]
Q_{32}&=&\left(EZ\frac{A}{D}\right)^\tra = E^\tra\frac{A}{D},\nn \\
&=&\frac{s}{s+t}\ \delta^{||}+\frac{p(\sh+t)}{2\Delta}\ iF
                                        +\frac{p\ch+t\sh}{2\Delta}\ \delta^\perp,
\nn \\ &  \equiv& s_3 \, \delta^{||}+\gamma_3 \, iF+\delta_3 \, \delta^\perp,
\nn \\ [10pt]
S&=&Z= \left(E-A\frac{A}{D}\right)^\tra=\delta^{||}+\frac{\al}{4\Delta}\ iF+\frac{\beta}{4\Delta}\ \delta^\perp,
\nn \\ &\equiv & r_1 \, \delta^{||}+\al_1 \, iF+\beta_1 \, \delta^\perp,
\nn \\ [10pt]
T&=&\left(EZ\right)^\tra=1+A^\tra\frac{A}{D}= \delta^{||}+\frac{2pq}{4\Delta}\ iF+\frac{p^2+q^2}{4\Delta}\ \delta^\perp,
\nn \\ &  \equiv& s_2 \, \delta^{||}+\gamma_2 \, iF+\delta_2 \, \delta^\perp,
\nn \\ [10pt]
U&=&\left(\frac{A}{D}\right)^\tra=\frac{s}{s+t}\ \delta^{||}+\frac{tp}{2\Delta}\ iF
                                        +\frac{p+t\sh}{2\Delta}\ \delta^\perp,
 \nn \\ &\equiv& r_2 \, \delta^{||}+\al_2 \, iF+\beta_2 \, \delta^\perp,
\nn \\ [10pt]
V&=&1-\frac{A}{D}=\frac{t}{s+t}\ \delta^{||}+\frac{tp}{2\Delta}\ iF
                                        +\frac{t q}{2N}\ \delta^\perp,
\nn \\ &  \equiv& s_1 \, \delta^{||}+\gamma_1 \, iF+\delta_1 \, \delta^\perp,
\eea
where \Ref{Z}, \Ref{A/D} and \Ref{umr} were used. The notations
$r_i$ $\al_1$, $\beta_i$, $s_i$, $\gamma_i$ and $\delta_i$  are
introduced here for later use.

As for the dependence on $\u$ we made use of the fact that only
the fourth  component of $\u_\mu$ is nonzero such that relations
like $Z\u=S\u=T\u=\u$ and $R\u=0$ hold.

In order to continue and, for instance, to find the necessary
structures for  integration by parts, we divide the contributions
into 3 parts,
\begin{enumerate}
  \item $M^{1,1}+M^{1,2}+M^{2,1}+M^{2,2}$,
\item $M^{1,3}+M^{2,3}+M^{3,1}+M^{3,2}$
  \item $M^{3,3}+M^{\rm gh}$
\end{enumerate}
and consider them individually in the following subsections.
\subsection{Contribution from $M^{1,1}+M^{1,2}+M^{2,1}+M^{2,2}$}
We start  with  $M^{1,1}$.
First of all we note that the relation
\be\label{pi1}
-\left(\frac{\pa}{\pa s}-\frac{\pa}{\pa t}\right) \left(P_{\la\la}-2\frac{\u_\la\u_{\la'}}{(\u p)}\right)
    =2Z\frac{2iF}{D^\tra}
\ee
holds which can be verified by differentiation of \Ref{Z} and
\Ref{umr}. It should be mentioned that the term
$-2\frac{\u_\la\u_{\la'}}{(\u p)}$ in the left  hand side vanishes
under differentiation. It was added by hindsight in order to carry out the
integration by parts and to derive a convenient representation for
the form factor.

Eq.\Ref{pi1} allows us to represent the contribution of $M^{1,1}$
to  the polarization tensor \Ref{Pi2} in the form (up to the sum
over $N$)
\be\label{}
\int dsdt \ \Theta_T(l^2,h^2+2iF)
    \left(-\left(\frac{\pa}{\pa s}-\frac{\pa}{\pa t}\right) \left(P-2\frac{\u_\la\u_{\la'}}{(\u p)}\right) \right) \tr E \ .
\ee
In this integral we temporarily change variables to $\la$ and $u$
according to $s=\la u$, $t=\la(1-u)$ and with
$\left(\frac{\pa}{\pa s}-\frac{\pa}{\pa
t}\right)=\frac{1}{\la}\frac{\pa}{\pa u}$ we integrate the
derivative with respect to $u$ by parts,
\bea\label{PI1}
&&\int_0^\infty d\la \la \int_0^1 du \ \Theta_T(l^2,h^2+2iF)
        \left(-\frac{1}{\la}\frac{\pa}{\pa u}
        \left(P-2\frac{\u_\la\u_{\la'}}{(\u p)}\right)
         \right) \tr E \nn\\
&=& \int_0^\infty d\la \la \  \left[ -  \Theta_T(l^2,h^2+2iF)
    \left(P-2\frac{\u_\la\u_{\la'}}{(\u p)}\right)
     \, \tr E \right] \Bigg|_{u=0}^1
\nn\\ && +\int dsdt \ \Theta_T(l^2,h^2+2iF)\Bigg\{
\left(P-2\frac{\u_\la\u_{\la'}}{(\u p)}\right)
\left(\widetilde{\left(pPp\right)}-2(\u p)\right)\ \tr E
\nn \\ &&~~~~~~~~~~~~~~~~~~~~~~~~~~~~~~~~~~~~~~~~~~~~~~~~~~~~~~~~~~~+4\sh P\Bigg\} \ ,
\eea
where
\be\label{pPp}{\left(pPp\right)}=
-\left(\frac{\pa}{\pa s}-\frac{\pa}{\pa t}\right)H-\frac{1}{2\Delta}\left(\frac{\pa}{\pa s}-\frac{\pa}{\pa t}\right)\Delta
\ee
follows from \Ref{H}, \Ref{N} and \Ref{PQetc} and
\be\label{diffT}
-\left(\frac{\pa}{\pa s}-\frac{\pa}{\pa t}\right)\Theta_T=\Theta_T
\left(\widetilde{\left(pPp\right)}-2(\u p)\right)
\ee
from \Ref{ThetaT}. We have to take into account that we
differentiate a $\Theta$, which  depends on $h^2+2iF$, so that we
get in place of \Ref{pPp}
\be\label{wtilde}
\widetilde{\left(pPp\right)}\equiv \left(pPp\right)_{\big|_{h^2\to h^2+2iF}} \ .
\ee
In such expressions we shall use the tilde as notation for this
substitution in the  following several times.  Further we used
$\tr E=2(1+\ch)$.

As a result from the integration by parts we get  surface
contributions which have the general form
\be\label{sfgen}M_{\rm surface}=\int_0^\infty d\la \la  \ \Theta_T(l^2,h^2+2iF) \sum_{i,j} M_{\rm sf}^{i,j}\Bigg|_{u=0}^1,
\ee
The surface contribution from $M^{1,1}$ is then
\be\label{sf11} M_{\rm sf}^{1,1}=-\left(P_{\la\la'}-2\frac{\u_\la\u_{\la'}}{(\u p)}\right) \ tr E \ .
\ee

Taking the   contribution from the last line in \Ref{PI1} we
represent  up to the surface contribution $M^{1,1}$ in the form
\bea\label{11a}
M^{1,1}&=&
\left(\left(Pp-2\u\right)_\la \left(P^\tra p-2\u\right)_{\la'}
-\left(P_{\la\la'}-2\frac{\u_\la\u_{\la'}}{(\u p)}\right)
\left(\widetilde{\left(pPp\right)} -2(\u p)\right)\right)\ \tr E
\nn \\ &&   +4\sh \left(P_{\la\la'}-2\frac{\u_\la\u_{\la'}}{(\u p)}\right) \ .
\eea

Next we consider $M^{1,2}+M^{2,1}$. It is useful to rewrite these in the form
\bea\label{}
M^{1,2}+M^{2,1}&=&
        \left(Pp-2\u\right)_\la \left(Rp\right)_{\la'}+
        \left(R^\tra p\right)_\la \left(P^\tra p-2\u\right)_{\la'}
\nn \\ &&
        -\left(P_{\la\la'}-2\frac{\u_\la\u_{\la'}}{(\u p)}\right)
        \widetilde{\left(pR^\tra p\right)}
        -R^\tra_{\la\la'}
                \left(\widetilde{\left(pP p\right)}-2(\u p)\right)
\nn \\ &&   +\left(P_{\la\la'}-2\frac{\u_\la\u_{\la'}}{(\u p)}\right)
                    (-4\sh)
        +R^\tra_{\la\la'}
                \left(\widetilde{\left(pP p\right)}-2(\u p)\right).
\eea
In the last line we use $\widetilde{\left(pR^\tra
p\right)}=-4\sh$. In fact, by the next-to-last and the last lines
we added zero. Now we shall integrate by parts in the last term in
the last line. Using the same  procedure as before and, for
instance, Eq.\Ref{diffT},  we get the surface contribution
\be\label{sf12}M^{1,2}_{\rm sf}=-\Theta_T(l^2,h^2+2iF)  R^\tra_{\la\la'}
\ee
and using
\be\label{}\left(\frac{\pa}{\pa s}-\frac{\pa}{\pa t}\right) R^\tra_{\la\la'}=8\ch \ iF
\ee
we represent
\bea\label{1221a}M^{1,2}+M^{2,1}&=&
   \left(Pp-2\u\right)_\la \left(Rp\right)_{\la'}+
        \left(R^\tra p\right)_\la \left(P^\tra p-2\u\right)_{\la'}
\nn \\ &&
        -\left(P_{\la\la'}-2\frac{\u_\la\u_{\la'}}{(\u p)}\right)
        \widetilde{\left(pR^\tra p\right)}
        -R^\tra_{\la\la'}
                \left(\widetilde{\left(pP p\right)}-2(\u p)\right)
\nn \\ &&-4\sh \left(P_{\la\la'}-2\frac{\u_\la\u_{\la'}}{(\u p)}\right)+8\ch \ iF_{\la\la'}.
\eea

Finally we turn to $M^{2,2}$. Here we rewrite
\bea\label{ti}
\left(pSp\right)=\widetilde{\left(pSp\right)}+\left(S+S^\tra\right)iF\ ,\nn\\
\left(pT^\tra p\right)=\widetilde{\left(pT^\tra p\right)}+\left(T+T^\tra\right)iF\
\eea
and represent $M^{2,2}$ in the form
\bea\label{22a}
M^{2,2}&=&-4\left[
    \left(Sp\right)_\la \left(Tp\right)_{\la'}+
    \left(T^\tra p\right)_\la \left(S^\tra p\right)_{\la'}
    -S_{\la\la'}\widetilde{\left(pT^\tra p\right) } -T^\tra_{\la\la'}\widetilde{\left(pSp\right)}\right]
     \nn \\&& -8\ch \ iF \ .
\eea
The last line is the result of a number of cancellations.

We continue with the observation that the last lines in \Ref{11a},
\Ref{1221a}  and \Ref{22a} compensate each other so that we are
left with the corresponding first lines. These are written in a
form as considered in Appendix A so that we can apply directly the
equations \Ref{M14}, \Ref{M56}, \Ref{ff0} and for the temperature
part \Ref{dW1} and so on. The first line in Eq.\Ref{11a} has the
form as given by \Ref{A0} and \Ref{W0} resp. \Ref{A0T} and
\Ref{W0T} multiplied by $\tr E$, the first lines of
Eqs.\Ref{1221a} and \Ref{22a} match the structure of $W_3$ in
\Ref{W} resp. \Ref{WT}. The parameters $r_i$ $\al_1$, $\beta_i$,
$s_i$, $\gamma_i$ and $\delta_i$ are given in Eq.\Ref{PQetc}.

In this way we get from this subsection the following contribution to the form factors, which we denote by $M^a_i$:
\be\label{gr1}
\begin{array}{rcll}
M^a_1=&-\left(\frac{s-t}{s+t}\right)^2 2(1+\ch)&&+8,\\ [8pt]
M^a_2=&\frac{(tp)^2-(p/2-t^2)^2}{\Delta^2} \ 2(1+\ch)&-8\sh\frac{tp}{\Delta}&+8\ch ,\\ [8pt]
M^a_3=&\frac{s-t}{s+t}\frac{t^2-p/2}{\Delta} \ 2(1+\ch)&&+\frac{\beta+p^2
                            +q^2}{\Delta},\\ [8pt]
M^a_4=&-\frac{s-t}{s+t}\frac{tp}{\Delta} \ 2(1+\ ch)&+4\sh\frac{s-t}{s+t}&+\frac{\al-2pq}{\Delta}  \\ [8pt]
&(11)&(12)+(21)&(22)
\end{array}
\ee
where in the last line the origin of the contribution is
indicated. We note that  there are no contributions to $M_5$,
$M_{6a}$ or $M_{6b}$.

For the temperature induced contributions we get with Eq.\Ref{W0T} with $\mu=2\frac{iN}{2(s+t)T}$
\bea\label{gr1T}
M^a_7&=&-2\frac{iN}{2(s+t)T}\ \frac{r_3}{(u p)}\ \tr E,\quad
M^a_7=-2\frac{iN}{2(s+t)T}\ \frac{\beta_3}{(u p)}\ \tr E,\nn \\
M^a_9=-M^a_{(*)}&=&2\frac{iN}{2(s+t)T}\ \al_3\ \tr E \ .
\eea

\subsection{Contribution from $M^{1,3}+M^{2,3}+M^{3,1}+M^{3,2}$}
In this subsection we employ the subsplitting \Ref{vf2}. First we
consider  $M^{3_1,1}+M^{3_1,2}$ from \Ref{M3} and do a reordering
using \Ref{PQetc},
\bea\label{M3_1}
M^{3_1,1}+M^{3_1,2}&=&
    2\left(Sp\right)_{\la}\left(Vp-\u\right)_{\la'}
    -2S_{\la\la'}\left(\left(pV^\tra p\right)-\u p\right)
                                -2\left(V+V^\tra\right)SiF  \\
&&\left(-Q_{31}^\tra p+\u\right)_{\la}\left(P^\tra p-2\u\right)_{\la'}
        +2\left(V+V^\tra\right)SiF-ZE^\tra\frac{2iF}{D^\tra}-2iFZ\frac{2itF}{D^\tra}
\nn
\eea
The reason for the reordering is that the last three terms in the second line collect into a derivative,
\be\label{} 2\left(V+V^\tra\right)SiF-ZE^\tra\frac{2iF}{D^\tra}-2iFZ\frac{2itF}{D^\tra}
=\left(\frac{\pa}{\pa s}-\frac{\pa}{\pa t}\right)
    \left(Q_{31}^\tra -\frac{\u_\la \u_{\la'}}{(\u p)}\right)\ ,
\ee
a relation which can be checked by explicit taking the
derivatives.  For instance, the term $-\frac{\u_\la \u_{\la'}}{(\u
p)}$ vanishes under differentiation. It was added by hindsight. In
a very similar way we rewrite $M^{1,3_1}+M^{2,3_1}$ in the form
\bea\label{}
M^{1,3_1}+M^{2,3_1}&=&
2\left(V^\tra p-\u\right)_{\la}\left(S^\tra p\right)_{\la'}
            -2S_{\la\la'}\left(\left(p V^\tra p\right)-\u p\right)
                                                -2\left(V+V^\tra\right)SiF
                                                \nn \\ &&
+\left(Pp-2\u\right)_{\la}\left(-Q_{31}p+\u\right)_{\la'}
       +\left(\frac{\pa}{\pa s}-\frac{\pa}{\pa t}\right) \left(Q_{31}^\tra -\frac{\u_\la \u_{\la'}}{(\u p)}\right)\  \ .
\eea
Now we integrate by parts in the same way as before. The surface contribution is
\be\label{M2sf}M^2_{\rm sf}=2\Theta_T(l^2,h^2+2iF) \  \left(Q_{31}^\tra -\frac{\u_\la \u_{\la'}}{(\u p)}\right)\
\ee
and in the bulk part we get
\bea\label{}
&&M^{3_1,1}+M^{3_1,2}+M^{1,3_1}+M^{2,3_1}\nn \\
 &&~~~~~~~~~=2\Big\{     \left(Sp\right)_{\la}\left(Vp-\u\right)_{\la'}+
            \left(V^\tra p-\u\right)_{\la}\left(S^\tra p\right)_{\la'}
                    -2S_{\la\la'}
                    \left(\widetilde{\left(pV^\tra p\right)} -\u p\right) \Big\}
\nn \\ &&~~~~~~~~~~~~~~
-\Big\{ \left(Pp-2\u\right)_{\la}\left(Q_{31}p-\u\right)_{\la'}
 \\ && ~~~~~~~~~~~~~~~~~~~~~~
        +\left(Q_{31}^\tra p-\u\right)_{\la}\left(P^\tra p-2\u\right)_{\la'}
        -2 \left(Q_{31}^\tra -\frac{\u_\la \u_{\la'}}{(\u p)}\right)\ \left(\widetilde{\left(pP p\right)} -\u p\right) \Big\} \nn \ ,
\eea
where also relations like \Ref{ti}  were used. We observe that the
two lines in the last formula have just the structure as one of
that given in Appendix A. For the upper line it is $W_2$ with
$\mu=0$ and $\nu=\frac{iN}{2(s+t)T}$ multiplied by an overall
factor of $2$ and for the lower two lines it is $W_1$ with
$\mu=2\frac{iN}{2(s+t)T}$ and $\nu=\frac{iN}{2(s+t)T}$ multiplied
by an overall factor of $(-1)$. The corresponding contributions to
the form factors can be obtained by means of \Ref{M14}, \Ref{M56},
\Ref{dW1} and \Ref{dW2}.

In a similar way the contributions from $M^{3_2,1}+M^{3_2,2}$ can be written as
\bea\label{}
M^{3_2,1}+M^{3_2,2}&=&
        2\left(T^\tra p\right)_{\la}\left(U^\tra p+\u\right)_{\la'}
        -2T^\tra_{\la\la'}\left(\left(p U p\right)+\u p\right)-2\left(U+U^\tra\right)T^\tra iF
\nn \\ &&
        +\left(Q_{32}^\tra p+\u\right)_{\la}\left(P^\tra p-2\u\right)_{\la'}
                +2\left(U+U^\tra\right)T^\tra iF - Z\frac{2iF}{D^\tra}
\eea
and and that from $M^{1,3_2}+M^{2,3_2}$ are
\bea\label{}
M^{1,3_2}+M^{2,3_2}
&=& 2\left(U p+\u\right)_{\la}\left(T p\right)_{\la'}
        -2T^\tra_{\la\la'}\left(\left(p U p\right)+\u p\right)-2\left(U+U^\tra\right)T^\tra iF
\nn \\ &&
        +\left(P p-2\u\right)_{\la}\left(Q_{32} p\right)_{\la'}
                +2\left(U+U^\tra\right)T^\tra iF - Z\frac{2iF}{D^\tra}.
\eea
In this case the derivative has the form
\be\label{} 2\left(U+U^\tra\right)T^\tra iF-Z \frac{2iF}{D^\tra}
=-\left(\frac{\pa}{\pa s}-\frac{\pa}{\pa t}\right)
\left(Q_{32}^\tra +\frac{\u_\la \u_{\la'}}{(\u p)}\right)\
\ee
where again $\frac{\u_\la \u_{\la'}}{(\u p)}$ is a term not
contributing here but added by hindsight. The resulting surface
contribution is
\be\label{M3sf}M^2_{\rm sf}=-2\Theta_T(l^2,h^2+2iF) \
\left(Q_{32}^\tra +\frac{\u_\la \u_{\la'}}{(\u p)}\right)
\ee
and the corresponding bulk parts after partial integration read
\bea\label{}
&&M^{3_2,1}+M^{3_2,2}+M^{1,3_2}+M^{2,3_2}
\nn\\&&  ~~~~~~~~~~~    =
        2\Big\{     \left(U p+\u\right)_{\la}\left(T p\right)_{\la'}
                  +  \left(T^\tra p\right)_{\la}\left(U^\tra p+\u\right)_{\la'}
                    -2T^\tra_{\la\la'}\left(\widetilde{\left(p U p\right)}+(\u p)\right)
            \Big\}
 \nn\\&&  ~~~~~~~~~~~~~~~~~~~
  +               \left(P p-2\u\right)_{\la}\left(Q_{32} p+\u\right)_{\la'}
                +    \left(Q_{32}^\tra p+\u\right)_{\la}\left(P^\tra p-2\u\right)_{\la'}
\nn \\ &&~~~~~~~~~~~~~~~~~~~~~~~~~~~~~~~~~~
                    -2\left(Q_{32}^\tra +\frac{\u_\la \u_{\la'}}{(\u p)}\right)\left(\widetilde{\left(p U p\right)}+(\u p)\right).
\eea
These have also the form as given in Appendix A, in both with
$W_1$.  In the upper line one has to take
$\mu=-\frac{iN}{2(s+t)T}$ and $\nu=0$ with an overall factor of 2
and in the lower lines $\mu=2\frac{iN}{2(s+t)T}$ and
$\nu=-\frac{iN}{2(s+t)T}$. Their contributions to the form factors
can be obtained by Eqs. \Ref{M14} and \Ref{M56} and \Ref{dW1} and
again we use the notations $r_i$ $\al_1$, $\beta_i$, $s_i$,
$\gamma_i$ and $\delta_i$ taken from \Ref{PQetc} and $r_4=r_3$,
$\al_4=\al_3$, $\beta_4=\beta_3$. Further we take into account the
numerical prefactors like 2 and -1, and we denote the
contributions to the form factors originating from this subsection
by $M^b_i$. These read
%
%
%

%
\bea\label{}
M^b_2&=&\frac{1}{2\Delta}\left(-\cosh ^2(2 s)+\left(2-4 t^2\right) \cosh (2 s)-12 t \sinh (2 s)\right),
\nn\\
M^b_3&=&\frac{1}{2(s+t)\Delta}\left(-2 t^3-10 s t^2-2 \left(t^2+s t+2\right) \cosh (2 s) t
\right.\nn\\&&~~~~~~~~~~~~~~~~\left.
-8 (s+t) \sinh (2 s) t+5 t-(s+t) \cosh ^2(2
   s)+s
    \right),
\nn\\
M^b_4&=&\frac{1}{2(s+t)\Delta}\left(4 t (t-s)-(s+t) \left(2 t^2+1\right) \sinh (2 s)
\right.\nn\\&&~~~~~~~~~~~~~~~~\left.
+\cosh (2 s) (4 (s-t) t+(s+t) \sinh (2 s))
    \right),
\nn\\
M^b_5&=&\frac{1}{2\Delta}\left((\cosh (2 s)-1) \left(2 t^2+\cosh (2 s)-1\right)
    \right),
\nn\\
M^b_{6a}&=&\frac{1}{2\Delta}\left(\left(-2 t^2+\cosh (2 s)-1\right) \sinh (2 s)
    \right),
\nn\\
M^b_{6b}&=&\frac{1}{2\Delta}\left(\left(-2 t^2+\cosh (2 s)-1\right) \sinh (2 s)
    \right).
\eea
We observe that $M^b_{6a}$ and $M^b_{6b}$ are equal. Also, we
remind that there was no contribution to $M^{6}$ from the
preceding subsection and there will be none from the next
subsection. As a consequence, the operators structures
$T^{(6a)}_{\la\la'}$ and $T^{(6b)}_{\la\la'}$ come with the same
form factors, hence these contributions collect into the structure
$T^{(6)}_{\la\la'}$, Eq.\Ref{T6}.

Finally we collect the temperature induced part. Its form factors
can be written in the form
\bea\label{gr2T}
M^b_7&=&\frac{iN}{2(s+t)T}\ \frac{1}{(u p)}\left(-2r_1+r_4+2s_4    +2s_2+r_3-2s_3\right) \ ,
\nn \\
M^b_8&=&\frac{iN}{2(s+t)T}\ \frac{1}{(u p)}\left(-2\beta_1+\beta_4+2\delta_4
                                                +2\delta_2+\beta_3-2\delta_3\right) \ ,
\nn \\
M^b_{10a}&=&\frac{iN}{2(s+t)T}\ \frac{1}{(u p)}\left(2r_1+r_4-2s_4   -2s_2+r_3+2s_3 \right) \ ,
\nn \\
M^b_{10b}&=&\frac{iN}{2(s+t)T}\ \frac{1}{(u p)}\left(2\beta_1+\beta_4-2\delta_4
                                                -2\delta_2+\beta_3+2\delta_3\right) \ ,
\nn \\
M^b_{9}&=&\frac{iN}{2(s+t)T}\ \left(2\al_1+\al_4+2\gamma_4+2\gamma_2-\al_3-2\gamma_4\right) \ ,
\nn \\
M^b_{(*)}&=&\frac{iN}{2(s+t)T}\ \left(2\al_4+2\al_3\right) \ .
\eea

\subsection{Contribution from $M^{3,3}+M^{\rm gh}$}
These contributions need a treatment to some extend different from
the preceding two subsections. First of all, we remove the factor
$Z$ introduced into $M^{3,3}+M^{\rm gh}$ in Eq.\Ref{M3} by means
of \Ref{ThetaZ} and we introduce a separate notation for the
corresponding contribution to the polarization tensor,
\be\label{Pi33}
\Pi^{33+{\rm gh}}_{\la\la'}=\sum_N \int dsdt\ \Theta_T(l^2,h^2) \ M^{33+{\rm gh}}_{\la\la'}
\ee
with
\bea\label{33+gh}
M^{3,3}_{\la\la'}+M^{\rm gh}_{\la\la'}&=&
    \delta_{\la\la} \left(
    \left(\left(    \left(1-\frac{A}{D}\right)p-\u\right)
                    \left(\left(1-\frac{A}{D}\right) p-\u\right)\right)
            -\tr E\frac{iF}{D^\tra} \ \right)
\nn \\  && + E_{\la\la'}  \left(
                    \left(\frac{A}{D}p+\u\right)\left( \frac{A}{D} \ p+\u\right)
            +\tr \frac{iF}{D^\tra} \ \right)  \ .
\eea
Now we use the statements that the relations
\be\label{pi332}
\Theta_T(l^2,h^2)
 \left(\left(    \left(1-\frac{A}{D}\right)p-\u\right)
                    \left(\left(1-\frac{A}{D}\right) p-\u\right)\right)
=-\frac{\pa}{\pa s} \ \Theta_T(l^2,h^2)
\ee
and
\be\label{pi331}
\Theta_T(l^2,h^2)   \left(
                    \left(\frac{A}{D}p+\u\right)\left( \frac{A}{D} \ p+\u\right)
            +\tr \frac{iF}{D^\tra} \ \right) =
-\frac{\pa}{\pa t} \ \Theta_T(l^2,h^2)
\ee
hold which can be verified by direct differentiation. In this way,
in \Ref{Pi33}  one integration can be carried out and we arrive at
\bea\label{sf33}
\Pi^{33+{\rm gh}}_{\la\la'}&=&
    -\sum_N \int_0^\infty ds\ E_{\la\la'}\Theta_T(l^2,h^2)\bigg|_{t=0}^{t=\infty}
    -\sum_N \int_0^\infty dt\ \delta_{\la\la'}\Theta_T(l^2,h^2)\bigg|_{s=0}^{s=\infty}.
\eea
%
\subsection{Tadpole and surface contributions}
In this subsection we collect the contributions resulting from the
tadpole  graphs given by Eq.\Ref{tadpol} and the surface
contributions which appeared in the preceding subsections.

The tadpole contributions can be calculated easily since they are
special cases of the basic loop contribution for $s=0$ collapsing
the line of the charged gluon and keeping the line of the neutral
gluon and for $t=0$ which the lines interchanged. All other rules
remain valid so that these contributions can be written down
easily,
\bea\label{tp4}
\Pi^{\rm tadpol}_{\la\la'}&=&
\sum_N \int dt \ \Theta_T(l^2,h^2) \left(-\delta_{\la\la'}+4 \delta_{\la\la}\right)_{\Big|_{s=0}}
\\ &&
+\sum_N \int ds \ \Theta_T(l^2,h^2) \left( \tr E \ \delta_{\la\la'}
-4\sh iF
-\left(\delta^{||}_{\la\la'}+iF\sh+\delta_{\la\la'}^\perp \right)
\right)_{\Big|_{t=0}}  \ .  \nn
\eea
Now we collect the surface terms. A part of them has the form \Ref{sfgen},
\be\label{sfgenl}
M_{\la\la'}^{\rm surface}=
\sum_N \int_0^\infty d\la  \ \Theta_T(l^2,h^2+2iF)  M_{\la\la'}^{\rm sf}\Bigg|_{u=0}^1
\ee
with contributions to $M_{\la\la'}^{\rm sf}$ from Eqs.\Ref{sf11}, \Ref{sf12}, \Ref{M2sf} and \Ref{M3sf},
\be\label{Msf}
M_{\la\la'}^{\rm sf}=
-\left(P_{\la\la'}-2\frac{\u_\la\u_{\la'}}{(\u p)}\right) \ \tr E
-R^\tra_{\la\la'}
+2{Q_{31}^\tra}_{\la\la'}-2{Q_{32}^\tra}_{\la\la'} \ .
\ee
which can be rewritten in the form
\bea \label{sf4a}
\Pi^{\rm surface}_{\la\la'}=
\sum_N \int ds \ \Theta_T(l^2,h^2)  {M_{\rm sf}}_{\big|_{t=0}}
-\sum_N \int dt \ \Theta_T(l^2,h^2){M_{\rm sf}}_{\big|_{s=0}}  \ .
\eea
Now it can be checked that when adding \Ref{tp4} and \Ref{sf4a}
and  doing the obvious simplification all contributions cancel
except for that which are proportional to $u_\la u_{\la'}$, i.e.,
to $T^{\rm D}_{\la\la'}$, Eq.\Ref{TD}. These collect into
$\Pi^{\rm D}$ defined in Eq.\Ref{expp},
\be\label{sf5}
\Pi^{\rm D}=
\sum_N \int ds \ \frac{iN}{2s}\  \Theta_T(l^2,h^2)  \left(-4+2\tr E\right)_{\big|_{t=0}}
-\sum_N \int dt \ \frac{iN}{2t}\ \Theta_T(l^2,h^2) \left(4-2\tr E\right)_{\big|_{s=0}} \ .
\ee
Now we need from \Ref{Theta2} and \Ref{ThetaT}
\bea\label{Th0}
{\Theta_T}_{\big|_{s=0}}&=&\frac{\exp\left(-\frac{N^2}{4tT^2}\right)}{(4\pi)^2 t^2} \ ,
\nn \\
{\Theta_T}_{\big|_{t=0}}&=&
    \frac{s}{\sinh(s)} \
\frac{\exp\left(-\frac{N^2}{4sT^2}\right)}{(4\pi)^2 s^2} \
  \exp\left(\frac{iN}{T}p_4\right)
\eea
and after renaming the integration variables we obtain for the remaining contributions
\bea\label{PiD1}
\Pi^{\rm D}&=&
  \sum_N  \int_0^\infty\frac{d\la}{\la}\  \frac{iN}{2Tp_4} \ \left[-4{\Theta_T}_{\big|_{s=0}}+4{\Theta_T}_{\big|_{t=0}}  \right]
\\ &=&-\frac{4}{(4\pi)^2} \sum_N
\int_0^\infty\frac{d\la}{\la^3}\ \   \frac{iN}{2Tp_4} \
    \left(1-\frac{\la\cosh(2\la)}{\sinh(\la)} \
                            \exp\left(i\frac{Np_4}{T}\right) \right)    \
        \exp\left(-\frac{N^2}{4\la T^2}\right)  \ . \nn
\eea
Thus, in our representation, the only contribution coming from the surface terms and tadpoles is the form factor $\Pi^{D}$.
\subsection{Debye mass of charged gluons}
The expression for $\Pi^{\rm D}$ derived in the preceding subsection
can be a bit simplified by
writing as a sum over $N>0$ (note the contribution from $N=0$
vanishes),
\bea\label{PiD2}
\Pi^{\rm D}&=&
 -\frac{4}{(4\pi)^2} \sum_{N=1}^\infty
\int_0^\infty\frac{d\la}{\la^3}\ \   \frac{\sin\left(Np_4/T\right)}{Tp_4/N} \
   \frac{\la\cosh(2\la)}{\sinh(\la)} \
        \exp\left(-\frac{N^2}{4\la T^2}\right)  \ .
\eea
It is obvious that for external momenta obeying $p_4=2\pi lT$ ($l$
integer)  only the contribution from $l=0$ is nonzero and we
arrive at the Debye mass in the field presence $\Pi^{\rm
D}=-\delta_{l,0}m_{\rm D}^2$,

\bea\label{mD}
m_{\rm D}^2(B)&=&
 \frac{1}{4\pi^2} \sum_{N=1}^\infty
\int_0^\infty\frac{d\la}{\la^2}\ \   \left(\frac{N}{T}\right)^2
   \frac{\cosh(2\la)}{\sinh(\la)} \
        \exp\left(-\frac{N^2}{4\la T^2}\right)  \ .
\eea
This expression coincides with the Debye mass of the neutral
gluon, Eq.(123)  in \cite{BORDAG2006E} that may serve as a check
of the correctness of carried out calculations.

The integral over $\la$ is  divergent at the upper limit  due to
the tachyonic mode. In fact, in this contribution one has to
written the parametric representation from the very beginning on
an axis rotated by 90 degrees clockwise and in the now divergent
contributions rotate in anti-Wick direction as this was done in
Ref \cite{BORDAG2006E}. As a result, in the high temperature
limit, $B/T^2 << 1$, one obtains
\bea\label{mDBfin} m^2_D(B) &&= \frac{2}{3}T^2 \left[1 - 0.8859
\left(\frac{\sqrt{B}}{2T}\right) + 0.4775 \left(\frac{B^2}{16 T^4
}\right) \right.\nn \\
&&\left. - i~ 0.4775 \left(\frac{\sqrt{B}}{2T}\right) +
O\left(\frac{B^3}{T^6}\right)\right], \eea
where the numeric values of the coefficients  are substituted. We
see that the real  part in the magnetic background is smaller than without the field. The imaginary part is small because it
appears in the next-to-leading order $\sim \sqrt{B} T$.

\section{Conclusion}
Here we collect the contributions which were calculated  in the
preceding subsections. The form factors appearing in the
decomposition \Ref{expp} of the polarization tensor read
\be\label{99}
\Pi^i(l^2,h^2+2iF)=
\sum_N \int_0^\infty ds \int_0^\infty dt \
\Theta_T(l^2,h^2+2iF)\ M_i
\ee
with
\bea\label{M0}
M_1&=&\frac{4 (s+t)^2-2 (s-t)^2 \cosh (2 s)}{(s+t)^2}
\nn\\
M_2&=&\frac{1}{\Delta}\left(-2 t^2+4 \sinh (2 s) t+2 \cosh ^2(2 s)+\cosh (2 s) \left(4 t^2+2 \sinh (2 s) t-3\right)+1
\right)
\nn\\
M_3&=&\frac{1}{4(s+t)\Delta}\left(2 (s+5 t) \cosh ^2(2 s)+4 t \left(t^2+5 s t-2\right) \cosh (2 s)\right.\nn\\&&~~~~~~~~~~~~~~~~\left.
+2 (s+t) \left(2 t^2+8 \sinh (2
   s) t-1\right)
\right)
\nn\\
M_4&=&\frac{1}{2(s+t)\Delta}\left(4 (s-t) t \cosh ^2(2 s)+(4 (s-t) t+(s-7 t) \sinh (2 s)) \cosh (2 s)
\right.\nn\\&&~~~~~~~~~~~~~~~~\left.
+8 t (t-s)+\left(-2 t^3+14 s
   t^2+7 t-s\right) \sinh (2 s)
\right)
\nn\\
M_5&=&\frac{1}{2\Delta}\left((\cosh (2 s)-1) \left(2 t^2+\cosh (2 s)-1\right)
    \right),
\nn\\
M_{6}&=&\frac{1}{2\Delta}\left(\left(-2 t^2+\cosh (2 s)-1\right) \sinh (2 s)
    \right),\eea
with $\Delta$ given by Eq.\Ref{N}. These are the same expressions
as in  our paper \cite{Bordag:2005br} dealing with the zero
temperature case.

The new temperature induced contributions    from adding
\Ref{gr1T} and \Ref{gr2T} read
\bea\label{MT}
M_7&=&\frac{iN}{2(s+t)Tp_4}\
\frac{4 (s-t) \cosh (2 s)}{s+t} \ ,
\nn \\
M_8&=&\frac{iN}{2(s+t)Tp_4}\
\frac{4 \cosh ^2(2 s)-4 \left(2 t^2+1\right) \cosh (2 s)}{2\Delta}\ ,
  \\
M_9&=&\frac{iN}{2(s+t)T}\
4\frac{t \cosh ^2(2 s)+(t+\sinh (2 s)) \cosh (2 s)-2 t+\left(2 t^2-1\right) \sinh (2 s))}{\Delta} \ . \nn
\eea
Furthermore it holds $M_{10a}=M_{10b}=0$ and $M_{(*)}=-M_9$.  Here
a comment is in order.  The vanishing of $M_{10a}$ and $M_{10b}$
is a result of the calculations done here in one loop order. At
the moment it is not known whether there is some symmetry behind
and whether this persists in higher loops. Similar remarks apply
to the relation between $M_{9}$ and $M_{(*)}$. As a result, these
two form factors contribute proportional to the tensor structure
\be\label{9-*}
T^9_{\la\la'}-T^{(*)}_{\la\la'}=u_\la id_{\la'}-id_\la u_{\la'}+iF(u p)-\frac{u_\la u_{\la'}}{(u p)}.
\ee

In addition to \Ref{99} with the form factors \Ref{M0} and
\Ref{MT} we also have the contribution from the Debye mass
\Ref{PiD2} which with \Ref{expp} and \Ref{TD} can be written in
the form, \be \Pi^{\rm D} u_\mu u_\nu. \label{NoT} \ee

It should be mentioned that  all the contributions \Ref{99}, and,
in particular, this one, are valid   off the shell of the
Matsubara values for the external momentum, i.e., for arbitrary
$p_4$. In this case, of course, even the weak transversality does
not hold and Eq.\Ref{NoT} is just the contribution on which this
is realized. For $\p_4=2\pi T l$ with integer $l$, from
Eq.\Ref{PiD2} $\Pi^{\rm D}= - \delta_{l,0}m_{\rm D}^2(B)$ follows
and, as discussed in section 2, the transversality holds.

Finally we  note  that, like in the case of the neutral
polarization tensor  in \cite{BORDAG2006E}, the form factors of
the charged one contain
 imaginary parts.  These result from the tachyonic mode and  must
  be treated in the same way as in \cite{BORDAG2006E}. An
  example is Eq. \Ref{mDBfin}.

One   obvious application of the results obtained in our
investigations is the resummation of perturbation series in the
field at high temperature. That can be done by means of the
solution of the Schwinger-Dyson equations for two-point Green's
functions, where the derived
 tensor structures with arbitrary form factors  have to
be used as the input propagators. Whether or not the spectrum of
gluons becomes stable due to the gluon magnetic mass, which is
generated in the field at high temperature in this so-called
super-daisy resummation, is an interesting problem for future
investigation.

\section*{Acknowledgements}
One of us (V.S.) was  supported by DFG by the Grant  No 436 UKR
17/24/05. He also thanks the Institute for Theoretical Physics of
Leipzig University for kind hospitality.

\section*{Appendix }
In this appendix we collect formulas which are necessary to
restore the form factors from the   expressions we are getting in
the main text for the parts of the polarization tensor under the
signs of the parametric integrals. In order not to overload the
notations we start from the $T=0$ case. A typical expression which
appears from the calculation of the polarization tensor in the
subsections 2.1 and 2.2  has the form
\be\label{A1}M=     
    \left(Xp\right)_{\la} \left(Yp\right)_{\la'}+
    \left(Y^\tra p\right)_{\la}  \left(X^\tra p\right)_{\la'}+ \ W_{\la\la'}
                ,
\ee
%
with
\bea\label{XYW}
    X&=&r \ \delta^{||}+\al \ iF+\beta \ \delta^\perp,\nn\\
    Y&=&s \ \delta^{||}+\gamma \ iF+\delta \ \delta^\perp,\nn\\
    W&=&a \ \delta^{||}+c \ iF+b         \ \delta^\perp.
\eea
We note that the parameters  in \Ref{XYW} has to fulfill certain
relations if $M$ shall obey the weak transversality condition
\Ref{wtc}. Irrespective of that, the first four form factors can
be restored.  Using \Ref{Ti} we rewrite \Ref{A1} in the form,
\bea\label{rest1}M&=&
        -2rs T^{(1)}-2(\al\gamma+\beta\delta)T^{(2)}
        -(r\delta+s\beta)T^{(3)}+(r\gamma-s\al)T^{(4)}
\nn\\
&&~~~~~~~~~~~+A \ \delta^{||}+C\ iF+B \ \delta^{\perp}  
\eea
with
\bea\label{ABC}
A&=&2rsl^2+(r\delta+s\beta)h^2+r\gamma-s\alpha+a \ ,
\nn\\
B&=&2\beta\delta h^2+(r\delta+s\beta)l^2+
            \al\delta-\beta\gamma+b \ ,
\nn\\
C&=&(s\al-r\gamma)l^2+(\al\delta-\beta\gamma)h^2+
            2\al\gamma+4\beta\delta+c\ .
\eea
From \Ref{rest1}, the first four form factors read
\be\begin{array}{rclrcl}\label{M14}
M_1&=&-2rs,&M_2&=&-2(\al\gamma+\beta\delta), \\
M_3&=&-(r\delta+s\beta),&M_4&=&r\gamma-s\al.
\end{array}
\ee

In the main text, the parts of the polarization tensor appear (partially after integrating by parts)  in  specific forms matching one of the following representations of $W$,
\bea\label{W}
W_1&=&-2Y^\tra_{\la\la'}\widetilde{\left(pXp\right)}   \ , \nn \\ [6pt]
W_2&=&-2X_{\la\la'}\widetilde{\left(pY^\tra p\right)}\ ,\nn \\
W_3&=&\frac12\left(W_1+W_2\right).
\eea
First we consider $W_1$ and use $\widetilde{\left(pXp\right)}=rl^2-\al+\beta(h^2+2iF)$ to obtain
\bea\label{Wi}
W_1&=&-2
        \left(s \, \delta^{||} -\gamma \, iF+\delta \, \delta^{\perp}\right)
        \left(rl^2  -\al +\beta(h^2+2iF) \right) ,\nn \\
&=&-2s\Big(
        \left(rl^2  -\al +\beta h^2 \right) \, \delta^{||} 
            +\left(-\gamma\left(rl^2  -\al +\beta h^2 \right)
                                            +\beta \delta\right) iF\nn \\ &&         ~~~~~~       +\left(\delta\left(rl^2  -\al +\beta h^2
                                       \right)-\beta\gamma\right)\delta^\perp
                \Big),
\eea
from which we identify the corresponding expressions for $a$, $b$ and $c$ in the second line in Eq.\Ref{XYW}. These we insert into $A$, $B$ and $C$ in \Ref{ABC},
\bea\label{W1ABC}
A&=&(r\delta-s\beta)h^2+r\gamma+s\al \ , \nn\\
B&=&-(r\delta-s\beta)l^2+3(\al\delta+\beta\gamma)\ , \nn\\
C&=&(r\gamma+s\al)l^2+(\al\delta+\beta\gamma)h^2 \ .
\eea
Comparison with \Ref{Ti} shows that these contributions collect
just into form factors such that $M$, \Ref{A1}, with $W_1$ from
\Ref{Wi} is  weak transversal. From \Ref{W1ABC} and \Ref{Ti} the
remaining form factors follow,
\be\label{M5}
M_5=r\delta-s\beta, \qquad M_{6a}=r\gamma+s\al, \qquad M_{6b}=\al\delta+\beta\gamma.
\ee
The same procedure can be repeated for $W_2$. The result is just
that the form factors $M_5$, $M_{6a}$ and $M_{6b}$ are the same
but with reversed sign. Finally, for $W_3$, they just compensate
each other. Collected together, these results read
\be\label{M56}
\begin{array}{c|cccc}
& M_5 &M_{6a}&M_{6b}\\\hline \\ [-10pt]
W_1 &   r\delta-s\beta  & r\gamma+s\al  &   \al\delta+\beta\gamma \\
W_2 &   -(r\delta-s\beta)  & -(r\gamma+s\al)  &   -(\al\delta+\beta\gamma) \\
W_3 &   0                   &       0           &       0                      &.
\end{array}
\ee

Also we need the special case when $Y^\tra =X$. In that case there
is only one  expression for $W$ and we have to consider (this is
$M$ from \Ref{A1} divided by two)
\be\label{A0}M=     
    \left(Xp\right)_{\la} \left(X^\tra p\right)_{\la'}+ \ W_{\la\la'}
                ,
\ee
with
\bea\label{W0}
W&=&-X_{\la\la'}\widetilde{\left(pXp\right)}   \ ,
\eea
In this case the nonzero form factors are
\be  \label{ff0}
M_1=-r^2,\quad M_2=al^2-\beta^2,\quad M_3=-r\beta,\quad M_4=-r\al \ .
\ee

Now we generalize these formulas to finite temperature. The basic expressions which appear in the subsections above have the form
\be\label{AT}M=     
    \left(Xp-\mu u\right)_{\la} \left(Yp-\nu u\right)_{\la'}+
    \left(Y^\tra p-\nu u\right)_{\la}  \left(X^\tra p-\mu u\right)_{\la'}+ \ W_{\la\la'}
                ,
\ee
Here $\mu$ and $\nu$ are numbers and for $W$ again 3 types of expressions appear,
\bea\label{WT}
W_1&=&
-2\left(Y^\tra_{\la\la'}-\nu\frac{u_\la u_{\la'}}{(up)}  \right)
\left(\widetilde{\left(pXp\right)} -\mu (up)\right)  \ ,
 \nn \\ [6pt]
W_2&=&-2\left(X_{\la\la'}-\mu\frac{u_\la u_{\la'}}{(up)}  \right)
\left(\widetilde{\left(pY^\tra p\right)} -\nu (up)\right)\ ,\nn \\
W_3&=&\frac12\left(W_1+W_2\right).
\eea
Let us  consider the contributions $\Delta W$ which are new as
compared to Eq.\Ref{A1}.  We start from the easiest part, namely
that proportional to $\mu\nu$. As easily can be seen from \Ref{AT}
and \Ref{WT} they cancel so that only contributions linear in
$\mu$ and linear in $\nu$ remain. For $W_1$ these read
\bea\label{dW1}
\Delta W_1&=&
-\nu\left[  \frac{r}{(up)}\left(T^{(7)}+T^{(10a)}\right)+
            \frac{\beta}{(up)}\left(T^{(8)}+T^{(10b)}\right)
                -\alpha\left(T^{(9)}-2T^{(*)}\right)\right]
\nn\\&&
-\mu\left[  \frac{s}{(up)}\left(T^{(7)}-T^{(10a)}\right)+
            \frac{\delta}{(up)}\left(T^{(8)}-T^{(10b)}\right)
                +\gamma T^{(9)} \right]\ . \eea
Here we used in addition to \Ref{Tt} the following formulas which are easy to verify,
\bea\label{}
T^{(7)}+T^{(10a)}&=&
\left(u_{\la}l_{\la'}+l_{\la}u_{\la'}\right)  (up)
    -2u_{\la}u_{\la'} l^2 \ ,
\nn\\
T^{(7)}-T^{(10a)}&=&
\left(u_{\la}l_{\la'}+l_{\la}u_{\la'}\right)  (up)
    -2\delta^{||}_{\la\la'} (up)^2 \ ,
    \nn\\
T^{(8)}+T^{(10b)}&=&
\left(u_{\la}h_{\la'}+h_{\la}u_{\la'}\right)  (up)
    -2u_{\la}u_{\la'} h^2 \ ,
\nn\\
T^{(7)}-T^{(10b)}&=&
\left(u_{\la}h_{\la'}+h_{\la}u_{\la'}\right)  (up)
    -2\delta^{\perp}_{\la\la'} (up)^2 \
\eea
and
\bea\label{}
T^{(9)}-2T^{(*)}&=&u_\la id_{\la'}-id_\la u_{\la'}-\frac{2}{(up)}u_\la u_{\la'}
\eea
with
\bea\label{}
T^{(*)}&=&\frac{1}{(up)}\left(u_{\la}u_{\la'}+(up)^2iF\right)  \nn \\
&=&\frac{1}{(up)}\ \frac{1}{p^2+2iF}\left((up)^2T^{(6)}-T^{(10)}\right) \ .
\eea
In a similar way we get
\bea\label{dW2}
\Delta W_2&=&
-\nu\left[  \frac{r}{(up)}\left(T^{(7)}-T^{(10a)}\right)+
            \frac{\beta}{(up)}\left(T^{(8)}-T^{(10b)}\right)
                -\alpha T^{(9)} \right]
\\&&\nn
-\mu\left[  \frac{s}{(up)}\left(T^{(7)}+T^{(10a)}\right)+
            \frac{\delta}{(up)}\left(T^{(8)}+T^{(10b)}\right)
                +\gamma \left(T^{(9)}-2T^{(*)}\right) \right].
\eea
Taking the half sum of both the expression
\bea\label{dW3}
\Delta W_3&=&
-\nu\left[  \frac{r}{(up)}T^{(7)}+ \frac{\beta}{(up)}T^{(8)}
-\al\left(T^{(9)}-T^{(*)}\right) \right]
\nn \\&&
-\mu\left[  \frac{s}{(up)}T^{(7)}+ \frac{\delta}{(up)}T^{(8)}
+\gamma\left(T^{(9)}-T^{(*)}\right) \right]
\eea
emerges.

Finally we note the special case with $Y=X^\tra$, $\mu=\nu$ for which we have
\be\label{A1T}M=
     \left(Xp-\mu u\right)_{\la} \left(X^\tra p-\mu u\right)_{\la'}+ \ W_{\la\la'}
                               ,
\ee
with
\be\label{A0T}
W=-\left(X_{\la\la'}-\mu\frac{u_\mu u_\nu}{(up)}\right)\left(\widetilde{\left( pXp\right)}-\mu (up)\right) \ .
\ee
We obtain in addition to \Ref{ff0}
\bea\label{W0T}
\Delta W &=&-\mu\left[
\frac{r}{(up)}T^{(7)} + \frac{\beta}{(up)}T^{(8)}
-\al\left(T^{(9)}-T^{(*)}\right)
 \right] \ .
\eea

\end{document}